\documentclass[preprint,prb,superscriptaddress,floatfix]{revtex4-1}

\newcommand{\pagenumbaa}{1}

\usepackage{graphicx}
\usepackage{color}
\usepackage{bm}
\usepackage{amsmath,amssymb}
\usepackage{epstopdf}
\usepackage{array,mathtools,amssymb,booktabs,tabularx}
\AtBeginDocument{
\heavyrulewidth=.08em
\lightrulewidth=.05em
\cmidrulewidth=.03em
\belowrulesep=.65ex
\belowbottomsep=0pt
\aboverulesep=.4ex
\abovetopsep=0pt
\cmidrulesep=\doublerulesep
\cmidrulekern=.5em
\defaultaddspace=.5em
}
\begin{document}
%%%%%%%%%%%%%%%%%%%%%%%%%%%%%%%%%%%%%%%%%%%%%%%%%%%%%%%%%%%%%%%%%%%%
% Place your title here
%%%%%%%%%%%%%%%%%%%%%%%%%%%%%%%%%%%%%%%%%%%%%%%%%%%%%%%%%%%%%%%%%%%%
\title{One law to rule them all: \\ Stretched exponential master curve of capacity fade for Li-ion batteries}
%%%%%%%%%%%%%%%%%%%%%%%%%%%%%%%%%%%%%%%%%%%%%%%%%%%%%%%%%%%%%%%%%%%%
% Place names of authors here
%%%%%%%%%%%%%%%%%%%%%%%%%%%%%%%%%%%%%%%%%%%%%%%%%%%%%%%%%%%%%%%%%%%%
\author{Eduardo Cuervo-Reyes}
\email{eduardo.cuervoreyes@ch.abb.com}
\affiliation{ABB Switzerland Ltd., Segelhofstrasse 1K, CH-5405 Baden-D\"attwil}
\thanks{E. Cuervo-Reyes contributed with the theoretical modelling, data analysis and wrote the article. R. Fl\"{u}ckiger designed the (ageing and characterization) experiments and performed the measurement.}
\author{Reto Fl\"{u}ckiger}
\affiliation{ABB Switzerland Ltd., Segelhofstrasse 1K, CH-5405 Baden-D\"attwil}

%%%%%%%%%%%%%%%%%%%%%%%%%%%%%%%%%%%%%%%%%%%%%%%%%%%%%%%%%%%%%%%%%%%%
% Abstract
%%%%%%%%%%%%%%%%%%%%%%%%%%%%%%%%%%%%%%%%%%%%%%%%%%%%%%%%%%%%%%%%%%%%
\begin{abstract}
Li-ion batteries gradually lose their capacity with time and use; therefore, ageing forecasts are key to designs  of battery powered systems. So far, cell-type-specific studies without standardised testing practices have lead to a variety of ageing models in which generality, simplicity, and accuracy seem exclusive. Previous studies hint to an interplay of multiple mechanisms leading to capacity loss, which depend on cell chemistry and are affected by temperature, state of charge, and cycling rate. Here we show that, despite this complexity, the time dependence of the actual capacity follows a unique master curve, for several cell types aged under various different conditions. We discuss  the statistical origin of this common behaviour, and the testing practice required for the characterisation of a model. The master curve is a stretched exponential that describes many other phenomena in nature and is theoretically justified within a diffusion-to-traps depletion model. These findings provide a simple and broadly applicable framework for  accurate life-time  predictions.
\end{abstract}
%\pacs{pacs number pending}
\maketitle
\setcounter{page}{\pagenumbaa}
\thispagestyle{plain}

%%%%%%%%%%%%%%%%%%%%%%%%%%%%%%%%%%%%%%%%%%%%%%%%%%%%%%%%%%%%%%%%%%%%
% Main Text
%%%%%%%%%%%%%%%%%%%%%%%%%%%%%%%%%%%%%%%%%%%%%%%%%%%%%%%%%%%%%%%%%%%%
\section{Introduction}
Ageing of rechargeable Li-ion cells  is characterised by the loss of cycleable lithium, and internal resistance increase\cite{Goodenough}.  These unwanted effects take place continuously in time and are accelerated with battery use. The ageing process of a Li-ion cell is commonly divided in three  stages\cite{Pastor2017301}. Firstly, there is a fast formation stage, corresponding to the initial passivation of highly reactive interfaces; e.g., the solid electrode-electrolyte interface (SEI). During the second stage, the SEI continues to grow slowly and the cell exhibits a decelerated degradation until its capacity is down to about 80 {\%} of its initial value. At this point the cell is considered to have reached the end of its life (EOL) for use. In the third stage, the capacity fade often accelerates and the internal resistance increases more abruptly due to multiple failure mechanisms.

Lithium losses result from trapping (reactions), either involving the SEI growth or elsewhere in the cell \cite{Verma,Winter}. Understanding these processes is essential for battery-material design and to optimise battery use. Quantifying lithium losses is a very complex task due to the variety of electrochemical reactions and transport phenomena involved\cite{Verma,Winter,SEImodelingreview}. This complexity is enhanced by the dependence of those processes on the changing  temperature, state of charge,  and electric current during normal use. To date, many  models have been developed with the purpose of predicting the capacity fade as a function of operating conditions. Some of these are based on microscopic equations\cite{Micros1,Sqrt}, some are purely phenomenological\cite{Experim1,Experim2,Aachen}, and others fully based on data statistics\cite{Databased1,Databased2}. Thus far, none have achieved a broad acceptance, nor is there a proof of the general applicability of any.

Motivated by the need for more efficient battery modelling tools, we performed extensive ageing experiments on cells from four different manufacturers (referred here on as Type 1--4). By means of a testing protocol that allows us to control the stress rate, it was found that the capacity retention as a function of time follows a unique master curve for all four types, independent of the ageing conditions. We show here experimental evidence of this uniform behaviour for the first time.     The master curve equally applies to cells that were cycled, as to those stored at a constant state of charge.  We discuss a physical interpretation of this uniform behaviour, and its relation to a universal form of relaxation. We also propose a generalization of the master curve in order to account for variable stress rates and predict capacity fade under arbitrary load profiles.

The article is organised as follows. Experimental evidence of the uniform fading law for three cell types is presented in section \ref{Evidence}. In section \ref{Physics}, we discuss the physical origin of this behaviour. Firstly, in \ref{DTTM} we revise the diffusion-to-traps model \cite{Grassberger1982}. Then, in \ref{Corr} we derive an important  relation between the capacity fade and the ion-trap correlation, and we discuss the relevant time scales involved. The time invariance of experimental observations, the influence of cycling, and the memory due to ion-trap correlation are reconciled in  \ref{Invar}, and \ref{Cycling}.  Further evidence from a fourth cell type, aged under not fully constant conditions, is presented in section \ref{Interrupted}. Section  \ref{Variabletau}  is dedicated to the  generalization of the model for the prediction of capacity fade under variable load profiles. A qualitative summary of the findings is given in section \ref{Summary}. Experimental details and mathematical derivations, which could dilute the main discussion of the work, can be found in appendices \ref{Meth}, and \ref{Support}.

\section{Experimental evidence of a universal ageing law}\label{Evidence}
Our first encounter with a previously unnoticed uniformity happened during the analysis of 22 identical cells (of Type 1), stressed in pairs according to 11 different ageing programs described in section \ref{Exper}, Table \ref{SettingsCell1}. A few cells were subject to calendar ageing only, kept at constant state of charge and temperature, $T$. Only the diagnostic cycles were performed, at two week intervals. The other cells were cycled continuously, each  with  constant charge and discharge current, and at constant $T$. The trajectory of the state of charge, $s(t)$, during cycling was always controlled and updated after every diagnostic cycle, in a way that the width of the cycle stayed proportional to the  remaining capacity. This means that for a cell with 1 Ah nominal capacity, a depth of discharge  of 90{\%}  corresponds to 0.9 Ah charge throughput during one charge (or discharge) at the beginning of life,  but only to 0.72 Ah at the EOL. The idea behind this setup is to keep the amount of stress constant per cycle in order to unravel the intrinsic relation between survival capacity and cycle number. Requesting a certain amount of charge from a system with less capacity causes a larger stress.  Thus, one needs to scale down the charge throughput with the available capacity.

\begin{figure}[htb]
\includegraphics[width=0.9\linewidth]{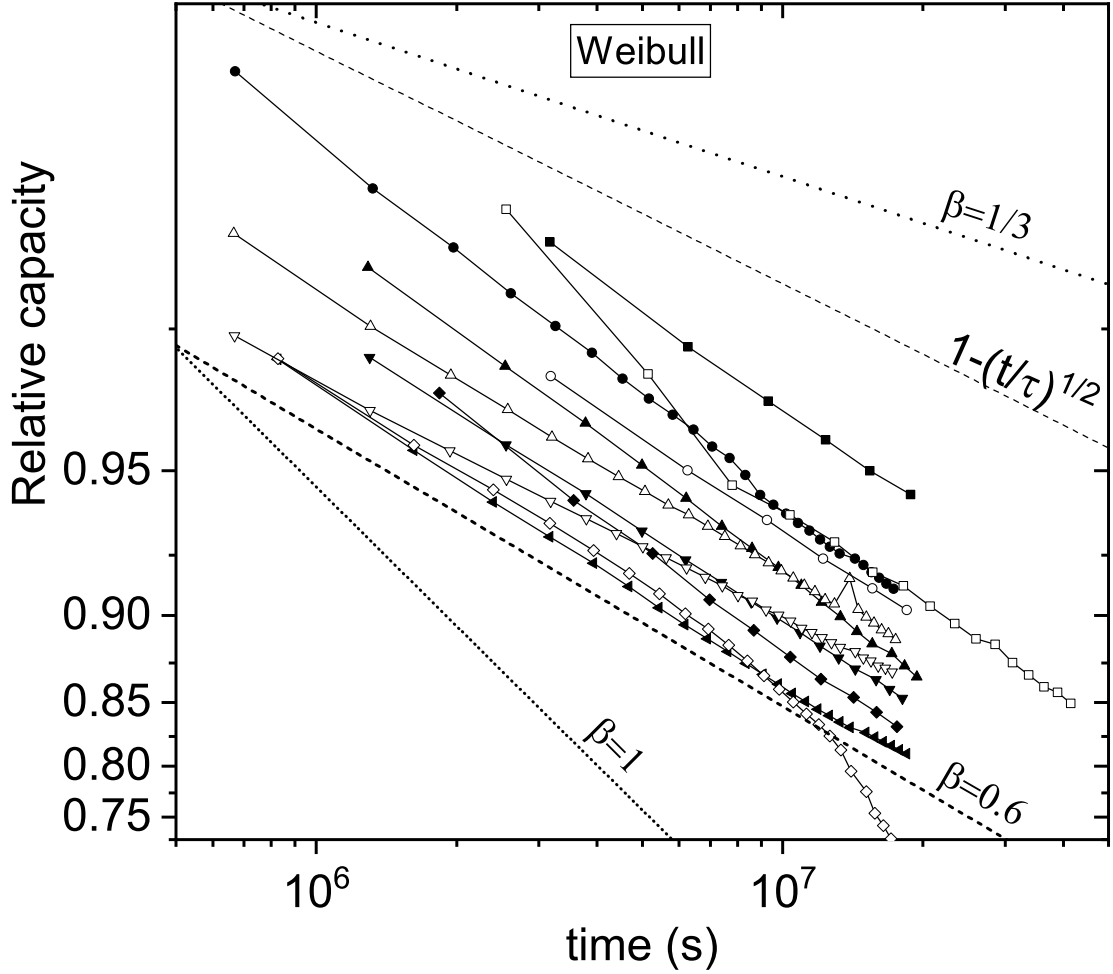}
\caption{Capacity retention vs time of 11 cells of Type 1 aged under different conditions. The ordinate axis is in $log(-ln(y))$ scale and the abscissa  in $log$-scale. In such a Weibull plot any stretched exponential $q=\exp(-[t/\tau]^{\beta})$ appears as a straight line with slope equal to $-\beta$. Solid lines between experimental points are guides to the eye. Model curves for $\beta=1/3$, $\beta=3/5$, and $\beta=1$,  as well as the commonly used model $q=1-\sqrt{t/\tau}$ are shown for comparison of the slopes.}
\label{fig:0}
\end{figure}

In FIG. \ref{fig:0} we display the fraction of the nominal capacity that remains as a function of time for the 11 different ageing programs. The choice of constant stress-rate unveiled  an interesting pattern, in that all trajectories are well represented by a stretched exponential (SE)
\begin{equation}\label{SER}
q=\exp\left(-\left[t/\tau\right]^{\beta}\right)
\end{equation} with exponent $\beta\approx 0.6$.  Note that in a Weibull plot any SE appears as a straight line with slope $-\beta$. Every trajectory is determined by its  time constant $\tau$. Curves with shorter $\tau$ appear towards the lower left of the graph; those with longer $\tau$, towards the upper right. In FIG. \ref{fig:0} we included model curves for SE with $\beta=1/3$, $\beta=3/5$, and $\beta=1$, as well as the well known model\cite{Sqrt} $q=1-\sqrt{t/\tau}$. The purpose is to compare the slopes, and the  model curves are arbitrarily located in the picture for better visibility. Note that, although the interval in which $q$ can be measured is rather narrow for a Weibull plot, an overall alignment of the data with a $-3/5$ slope is clearly visible. In FIG. \ref{fig:Meth1} section \ref{Support} (supplements), we show in more detail that Eq.(\ref{SER})  fits the data better than  $q=1-\sqrt{t/\tau}$. The latter underestimates the ageing when extrapolating from short-time experiments.

\begin{figure}[htb]
\includegraphics[width=0.9\linewidth]{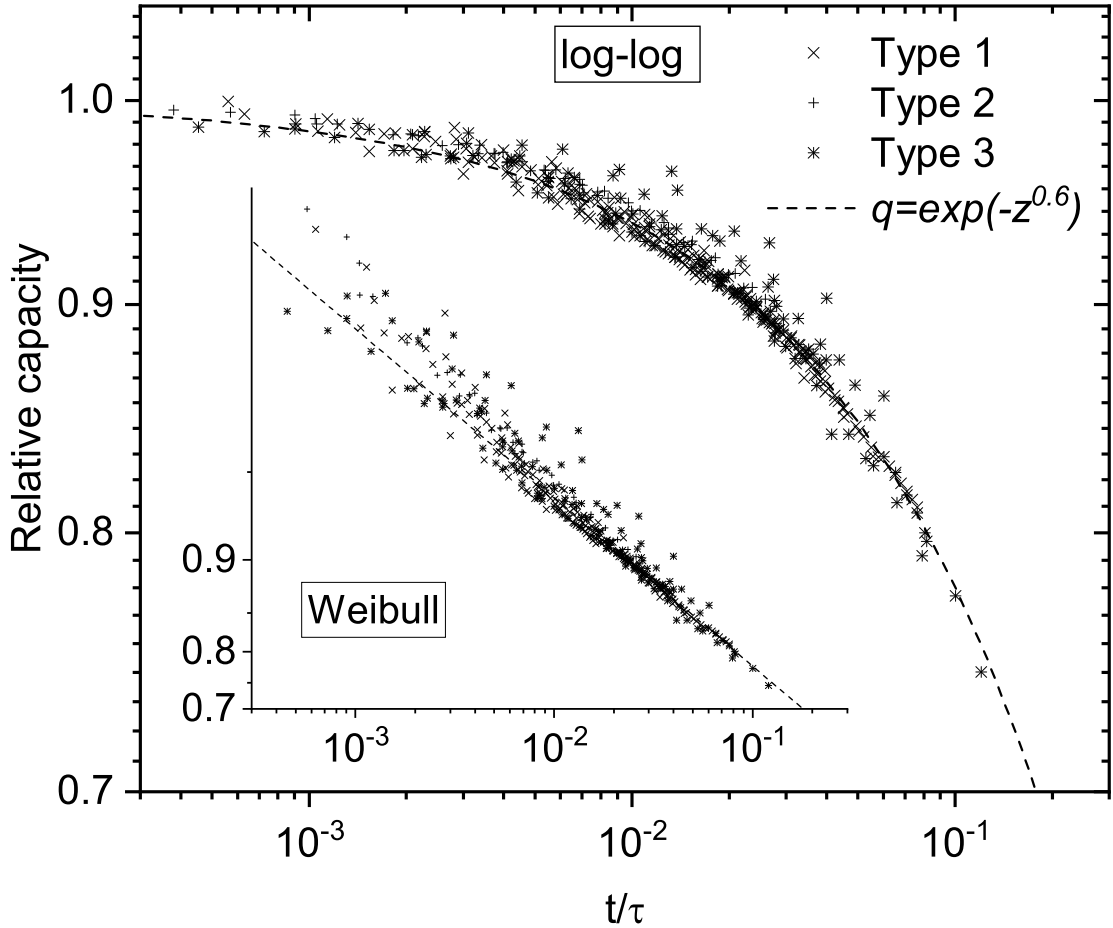}
\caption{Capacity retention vs time of 28 cells from three different manufacturers, aged under different conditions. For every trajectory, the time $t$ is given in units of the characteristic time constant $\tau$  determined via a fit with $y=\exp(-[t/\tau]^{0.6})$. The plot contains  over 400 data points in total. The other set of twin trajectories corresponding to (reproducibility tests with) other 28 cells has been excluded for clarity. The main panel uses $\log$ vs $\log$ scaling of the axes. The inset shows the Weibull scaling. Note that as a function of the dimensionless variable $t/\tau$, all 28 trajectories follow the master curve.}
\label{fig:1}
\end{figure}

Capacity measurements often show a considerable amount of scattering, and can be larger than 2 {\%} even after occasional failures are excluded. Thus, in order to better judge the statistical significance of the observed pattern, we redisplay the data introducing a scaled dimensionless abscissa $z=t/\tau$ ($\tau$ being the corresponding time constant for each trajectory).  If Eq.(\ref{SER}) is the correct model, all trajectories plotted with respect to the corresponding $t/\tau$  should collapse on a single master curve $q=\exp(-z^{0.6})$ that has no adjustable parameter. We took this test one step further, including the data of other 17 cells from two other manufacturers (Type 2 and Type 3). The ageing conditions for these 17 cells also adhere to regular cycling plans with constant stress per cycle, which are described in section \ref{Meth}, tables \ref{SettingsCell2} and \ref{SettingsCell3}.   The 28 trajectories are shown in FIG. \ref{fig:1}. In the main panel, the capacity retention vs time is plotted in a log-log scale. The inset displays the same data but with the Weibull scaling of the axis.  Both plots show a good agreement of the 405 data points with the master curve; the latter being represented with a dashed line. Due to the double log scaling of the $y$-axis in the Weibull plot, data at short times appear more scattered, which simply corresponds to the typical fluctuations of about 2{\%}. The Weibull plot highlights the good alignment with the slope $-3/5$. FIG. \ref{fig:1} confirms the statistical significance of the pattern. It is highly improbable that trajectories of 28 cells from three different brands, and aged under different conditions all fall on the same master curve due to pure casuality. This provides strong evidence of a common underlying dynamic governing the capacity fade of all cells studied here.

\section{Physical explanation}\label{Physics}

It turns out that Eq.(\ref{SER}) represents a recurrent pattern in nature. Eq.(\ref{SER}) has  been found to describe the conductivity relaxation in ion-, and electron-conductors close to the metal-insulator transition\cite{PhysRevLett.96.037403,PhysRevLett.76.1296,Kohl,NGAI1996,NgaiBook2017}; the magnetic relaxation in frustrated magnets\cite{PhysRevLett.97.047203}, spin glasses at the critical point\cite{PhysRevLett.72.1291,Ogielski}; the structural relaxation in glass forming liquids\cite{PhysRevLett.58.571,PhysRevLett.68.71}, polymers, supercooled liquids, and quasicrystals\cite{Phillips1996,Phillips2012}. Phenomena outside physical sciences such as  the probability of literature citation chains\cite{Trachenko2006},  human dynamics\cite{HumanDyn}, and the jamming transition\cite{PhysRevLett.99.060604} are described by this function as well.  This ubiquitous  behaviour is understood in terms of the Lifshitz-Kac-Luttinger  diffusion-to-traps depletion model\cite{Grassberger1982}, which we describe below. The exponent $\beta$ only takes a few discrete values depending on the dimensionality of the space and constitutes a unique example of magic numbers in classical physics\cite{Phillips1996,Mauro2012}.

\subsection{The diffusion-to-traps model}\label{DTTM}
The diffusion-to-traps model consists of particles diffusing in a $d$-dimensional space where traps are randomly   distributed. A particle that falls in a trap is considered lost. In the case of batteries, the particles are the Li-ions and the traps are the points where these ions can be irreversibly bound. The number of traps $N$ per volume $V$ is clearly an important parameter; but their random distribution plays an essential role too, as we show later. There are many ways in which $N$ trapping points can be randomly located in a volume $V$.  Let $S$ denote an arrangement of traps, $P(S)$ its probability, and $\rho_S(x)$ the corresponding trap density function, where $x$ is the position vector in the $d$-dimensional space. The probability density, $\rho(x,t|S)$, of finding an ion at position $x$ after time $t$,  given that traps are in configuration $S$,  is the solution of  the partial differential equation
\begin{eqnarray}
\partial_t\rho(x,t|S)&=&D\nabla^2\rho(x,t|S)\label{General}\\
&-&\int_V d^dx' \rho_S(x')K(x'-x)\rho(x,t|S)\nonumber
\end{eqnarray}
\noindent where $K(x'-x)$ is the trapping rate, $D$ is the diffusion coefficient,  $\partial_t$ is the partial derivative with respect to time, and $\nabla^2$ is the Laplace operator. The total survival probability $q(t)$ is the fraction of particles that remains un-trapped after a certain time.  It is obtained from the volume integral $q(t)=\int_V d^dx\, \bar{\rho}(x,t)$ of the  disorder-averaged density
\begin{equation}
\bar{\rho}(x,t)\equiv\langle\rho(x,t|S)\rangle=\sum_S \rho(x,t|S)P(S)\; , \label{avdens}
\end{equation}
\noindent with initial condition $\rho(x,0|S)=\delta(x)$. Although solving Eq.(\ref{General}) for every $S$ requires large-scale numerical simulations, the asymptotic short- and long-time forms of $q(t)$ can be worked around.

\subsubsection{Initial decay rate. Short-time behaviour}
The initial decay rate, $-\frac{1}{q}\frac{dq}{dt}\vert_{t=0}$, can be computed exactly giving
\begin{equation}
\tau_0^{-1}=n_S\int d^dx K(x)\,,
\end{equation}
\noindent where
\begin{equation}
n_S\equiv \langle\rho_S(x)\rangle=N/V
\end{equation}
\noindent is the global density of traps. Ions within the action range of traps are lost at a constant  rate $\tau_0^{-1}$. The same result is obtained for $t>0$ in the mean-field approximation. That means, if one neglects the correlation between $\rho(x,t|S)$, and $\rho_S(x)$ by imposing that $\langle\rho(x,t|S)\rho_S(x')\rangle=\langle\rho(x,t|S)\rangle\langle\rho_S(x')\rangle$,
the solution of  Eq.(\ref{General}) simplifies to

\begin{equation}
q(t)\approx \exp(-t/\tau_0^{})\, .\label{exp}
\end{equation}

\noindent  However, as time passes, the exact solution to Eq.(\ref{General})  gives higher survival probability density in regions with lower density of traps. This effect is discounted with the mean-field approximation.  Therefore Eq.(\ref{exp}) loses validity around a time $t_c<\tau_0$ and there must be a crossover to a new regime in which the global trapping rate decreases. Further ions can be trapped only after casually moving within trapping distance as a result of diffusion. Trapping becomes a diffusion-limited process.
\subsubsection{Long-time decay function}
In section \ref{Theor} we derive that $q(t)\ge A_d\exp\left(-[t/\tau]^{\beta}\right)$, where
\begin{equation}\label{be}
\beta=d/(d+2)
\end{equation}
\noindent and $A_d$, and $\tau$ are also constants. This can be understood with a simplified reasoning. Due to the randomness in the location of the traps, the probability that there are no traps within a distance $R$ to an ion can be derived from the Poisson distribution. This gives $P_0=\exp(-n_Sc_dR^d)$, where   $c_d=\frac{\pi^{d/2}}{\Gamma(d/2+1)}$, and $\Gamma(d/2+1)$ is the complete gamma function. The probability that the ion does not travel further than $R$ within the interval $t$ is calculated solving the diffusion equation. For long times, it is proportional to  $\exp(-\gamma_1 D t/R^2)$, where $\gamma_1$ is a number of order 1. This probability increases with $R$, while $P_0(R)$ decreases strongly as $R$ increases. The meaning is that large regions without traps are less probable than smaller ones, but in those large regions the ions survive longer times. The product   $\exp(-\gamma_1 Dt/R^2)P_0(R)$ strongly peaks at $R=R_0(t)\equiv \left(\frac{2\gamma_1D}{dc_dn_S}t\right)^{\frac{1}{d+2}}$  as shown in FIG. \ref{Rsignificance}. $R_0(t)$ is the typical distance travelled by a particle until it becomes trapped. In contrast to the case with homogeneous trap density $\rho_S(x)\equiv n_S$ (mean-field) where $R_0$ would be a constant proportional to $n_S^{-1/d}$, here $R_0$ increases with time. While the former  case would lead to an exponential decay, the time dependent $R_0(t)$ leads instead to a SE decay as in Eq.(\ref{SER}), with

\begin{equation}
\tau^{-1}=\frac{\pi\gamma_1 D n_S^{2/d}(1+2/d)^{1+2/d}}{ \left[\Gamma(1+d/2)\right]^{2/d}}\,,\label{tau}
\end{equation}

\noindent and $\beta$ given by Eq.(\ref{be}).  Eq.(\ref{SER}) is valid for any trapping rate $K(x-x')$, as long as the trapping points occupy a small fraction, $f$, of the total volume and are randomly distributed \cite{Grassberger1982}. As such, the situation described by Eq.(\ref{SER}) is very general. Many relaxation phenomena in nature  can be described as diffusing quasi-particles that decay at randomly located defects\cite{Phillips1996,Mauro2012}. The evidence from ageing of Li-ion cells presented in the previous section suggests that their capacity fade is governed by this same dynamic.   The experimentally found exponent $\beta\approx 0.6$  is in perfect agreement with the theoretical prediction of Eq.(\ref{be}), for the diffusion in a space with  ($d=3$) three dimensions. One can further verify for consistency, that $f$ is small. For all curves in Fig.(\ref{fig:0}) and Fig.(\ref{fig:Meth1}), $\tau >10^8\, \text{s}$. Considering the typical diffusion coefficients for Li-ions, $D<10^{-14}\,\text{m}^2\text{s}^{-1}$, such $\tau$ values correspond to trap densities $n_S< 10^{9}\,\text{m}^{-3}$, which compared with atomic densities $\sim 10^{29}\,\text{m}^{-3}$, implies $f< 10^{-20}\ll 1$ as required.   Nevertheless, there are a few points that should be clarified in order to fully understand the general applicability of  Eq.(\ref{SER}).

So far the model has been described as if one could follow $q(t)$ from an absolute time $t=0$ in which particles and traps were uncorrelated.  However, the level of correlation at the beginning of an experiment is unknown in practice. The model should account for the observed SE decay with arbitrary starting time $t_0$. Therefore, we must reformulate it (if possible) in a from that is compatible with the multiplication of probabilities, $q(t,0)=q(t,t_0^{})q(t_0^{},0)$. We should also clarify why the capacity fade of cycled batteries exhibits the same time dependence as that of stored ones. One would think that cycling erases any correlation by redistributing the particles.  In the following, we address these points by looking at analytical properties of $q(t)$, and taking into consideration the ion dynamics during cycling.

\subsection{Ion-trap density correlation and decay rate}\label{Corr}
The short- and long-time limits of $q(t)$ given by Eq.(\ref{exp}) and Eq.(\ref{SER}), respectively,  must be smoothly connected. The corresponding analytical function  is\cite{IonCond}
\begin{equation}
q(t,0)=\exp\left\{\frac{t_c}{\beta\tau_0^{}}\left[1-\left(1+\frac{t}{t_c}\right)^{\beta}\right]\right\}\, , \label{Analyt}
\end{equation}
\noindent with a crossover time $t_c=\left(\frac{\beta\tau_0^{}}{\tau^\beta}\right)^{1/(1-\beta)}$ that depends on the microscopic parameters as
\begin{equation}\label{tc}
t_c= \tau_0 \left(\frac{1}{1+d/2}\right)^{1+d/2}\left(\tau_0 D n_S^{2/d}\right)^{d/2}\, .
\end{equation}
To find the relation between the decay rate $\kappa(t)=-\frac{1}{q}\frac{dq}{dt}$ and particle-trap correlation, we  formally calculate $\frac{dq}{dt}$ using Eq.(\ref{General}), and compare it with the derivative of Eq.(\ref{Analyt}). Volume integration and disorder-average of  Eq.(\ref{General}) give
\begin{equation}\label{dqdt}
\frac{dq}{dt}=-\sum_S P(S)\int_V dx\int_V dx'\rho_S(x')\rho(x,t|S)K(x-x')\,.
\end{equation}
\noindent Since Eq.(\ref{Analyt}) is valid for any short-ranged $K(x-x')$, we can assume $K(x-x')=\Omega \delta(x-x')$ without losing generality. $\Omega$ is a constant with units of volume$/$time. This allows us to formally evaluate the right hand side of  Eq.(\ref{dqdt}), and to find that
\begin{equation}\label{dqdtsimplif}
\kappa(t)= n_S \Omega \left(1+C(t)\right)\, ,
\end{equation}
\noindent where
\begin{equation}\label{Defcorrel}
C(t)\equiv \frac{1}{q(t)n_S}\int_V dx \sum_S P(S)\rho_S(x)\rho(x,t|S)-1
\end{equation}

\noindent is the normalized correlation function. Eq.(\ref{dqdtsimplif}) provides the quantitative relation between particle-trap correlation and decay rate.  Substituting Eq.(\ref{Analyt}) in Eq.(\ref{dqdtsimplif}) one obtains for the particle-trap correlation \begin{equation}
 C(t) = \left(1+t/t_c\right)^{\beta-1}-1\, .
\end{equation}
At $t=0$,  $\kappa$ takes its maximum value $\kappa(0)\equiv\tau_0^{-1}=n_S\Omega$ and $C(0)=0$.  The decrease in the decay rate as time passes is mirrored by the development of negative correlations $-1< C(t)<0$, due to the lower particle density in regions with higher density of traps. $\kappa(t)$ decreases gradually without a characteristic cut-off time. For $t\gg t_c$,  it  follows a (scale-invariant) power law $\kappa(t)\propto(t_c/t)^{1-\beta}$.  The crossover time, $t_c$, depends on how fast particles move. In the hypothetical limit of infinitely fast diffusion $D\rightarrow \infty$, distances become irrelevant, $t_c^{}\rightarrow\infty$ and Eq.(\ref{Analyt}) reduces to Eq.(\ref{exp}) (equivalent to  the mean-field approximation, because correlations cannot develop). In the opposite extreme, $D\rightarrow 0$, the global decay rate vanishes as most particles take infinite time to reach a trap. For Li-ion batteries, $t_c^{}$  is much smaller than typical ageing time scales. Take for instance the ageing curves in Fig.(\ref{fig:0}) and Fig.(\ref{fig:Meth1}). We have found that  $\tau>10^8\,\text{s}$, which corresponds to  $f< 10^{-20}$. With $f$, and with the trapping rate at a trapping point ($\nu_0$), we can estimate $\tau_0^{-1}=\nu_0 f$ and  $t_c^{}$ according to Eq.(\ref{tc}). Since  $\nu_0$ correspond to molecular time scales $\nu_0\sim 10^3\,\text{THz}$, one finds that $\tau_0^{}\sim 10^5\,\text{s}$, and $t_c^{}\sim 1\,\text{s}$.

\subsection{Asymptotic time invariance}\label{Invar}
In order to generalise  Eq.(\ref{Analyt}) to  arbitrary initial time $t_0$,  the relative survival probability $q(t,t_0)$ is calculated by integrating $\kappa(t)$ between $t_0$ and $t=t_0+\Delta t$,
\begin{equation}\label{generalized}
q(t,t_0)=\exp\left\{-\int\displaylimits^{t}_{t_0}\frac{dz}{\tau_0}\frac{1}{\left(1+z/t_c\right)^{1-\beta}}\right\} \,.
\end{equation}
\noindent Rewriting  Eq.(\ref{generalized}) in terms of a new integration variable  $u=z-t_0$, it becomes
\begin{equation}\label{generalizedrel}
q(t,t_0)=\exp\left\{-\int\displaylimits^{\Delta t}_{0}\frac{du}{\tau_0^{'}}\frac{1}{\left(1+u/t'_c\right)^{1-\beta}}\right\} \, ,
\end{equation}
\noindent with $t_c'=t_c+t_0$, and $\tau_0^{'}=\tau_0^{}\left(1+t_0/t_c\right)^{1-\beta}$. This shows that the functional form of $q(t,t_0)$ is invariant to time shifts if one simultaneously scales the microscopic parameters ($t'_c$ for $t^{}_c$, and $\tau'_0$ for $\tau_0^{}$). A decay observed after $t=t_0$ will display the same functional form as when observed from $t=0$, but will have smaller decay rates and a longer crossover time. However, the most useful property is that $q(t,t_0)$ has a memoryless lower bound $q^{}_{\text{lb}}(t-t_0^{})\leq q(t,t_0)$,

\begin{eqnarray}
q^{}_{\text{lb}}(\Delta t)&=&\exp\left\{-\int\displaylimits^{\Delta t}_{0}\frac{du}{\tau_0^{'}}\left(\frac{t'_c}{u}\right)^{1-\beta}\right\}\label{longdT} \\
&=&\exp\left(-\left[\Delta t/\tau\right]^{\beta}\right) \label{SErel}\, ,
\end{eqnarray}

\noindent  because $(t'_c)^{1-\beta}/\tau_0^{'}=\ (t_c)^{1-\beta}/\tau_0^{}=\beta/\tau^{\beta}$ does not depend on $t_0$.  Furthermore, over long observation periods, $\Delta t\gg t_0+t_c$,  $q(t,t_0)\rightarrow q^{}_{\text{lb}}(\Delta t)$.

The analysis presented above can be summarised as follows. During a process of particles diffusing in a space with traps, the global trapping rate decreases with time, since the overlap of trap- and particle-density decreases faster than the total number of particles.  This process has memory, because the level of overlap (and correlation) depends on the time measured from the initialization moment, when particles were randomly distributed. The exact solution for the survival probability satisfies the multiplication of probabilities $q(t,0)=q(t,t_0)q(t_0^{},0)$, but in order to compute $q(t,t_0)$ exactly, one needs information that is generally unavailable. That is, one needs  to know the trapping rate  $1/\tau'_0$ and the crossover time $t'_c$ at the starting point $t_0^{}$. Fortunately, $q(t,t_0)$ has a time-invariant lower bound given by Eq.(\ref{SErel}). The lower bound requires a single parameter $\tau$ that is time invariant. This is compatible with the experimental reproducibility of the capacity fade. That is, batteries of a same type, aged under identical conditions, show fading curves with approximately the same $\tau$, regardless of their previous history. This also means that the parameter $\tau$ extracted from ageing experiments with a battery of any age can be used  to predict the capacity loss of older, and  newer batteries.  The lower bound does not satisfy exactly the multiplication of probabilities, but becomes asymptotically accurate for long time intervals.

\subsection{Effect of cycling}\label{Cycling}
The SE decay  has been theoretically explained within the framework of a diffusion-to-traps model,  which relies on  two basic assumptions: (i) short-range traps are randomly distributed, and (ii) particle movement is governed by the diffusion equation. The latter implies a linear  scaling  of the mean-square-displacement of the ions with time, $\text{MSD}\propto t$. Thus, whether cycling can change the functional form of the decay or not, reduces to whether it changes this scaling. The answer is found in impedance spectroscopy studies published elsewhere\cite{Macdonald2005}. The impedance of Li-ion batteries exhibits a frequency  dependence $Z(\omega)\propto (i\omega)^{-n}$ with $n\approx 1/2$, for frequencies in the range 1 mHz $<\omega< 1$ Hz.   Frequencies below 1 Hz correspond to  typical time scales for Li-ion intercalation (cycling) and the observed functional form of $Z(\omega)$, to a diffusion-limited transport. This implies that during cycling, ion transport is still dominated by density gradients and therefore, the scaling $\text{MSD}\propto t$ is preserved.  Cycled batteries may lose their capacity  faster, but will display the same functional dependence as stored batteries. This scenario is consistent with the experimental results presented in section \ref{Evidence}.

\section{Further evidence: irregular cycling}\label{Interrupted}

\begin{figure}[htb]
\centerline{\includegraphics[width=0.8\linewidth]{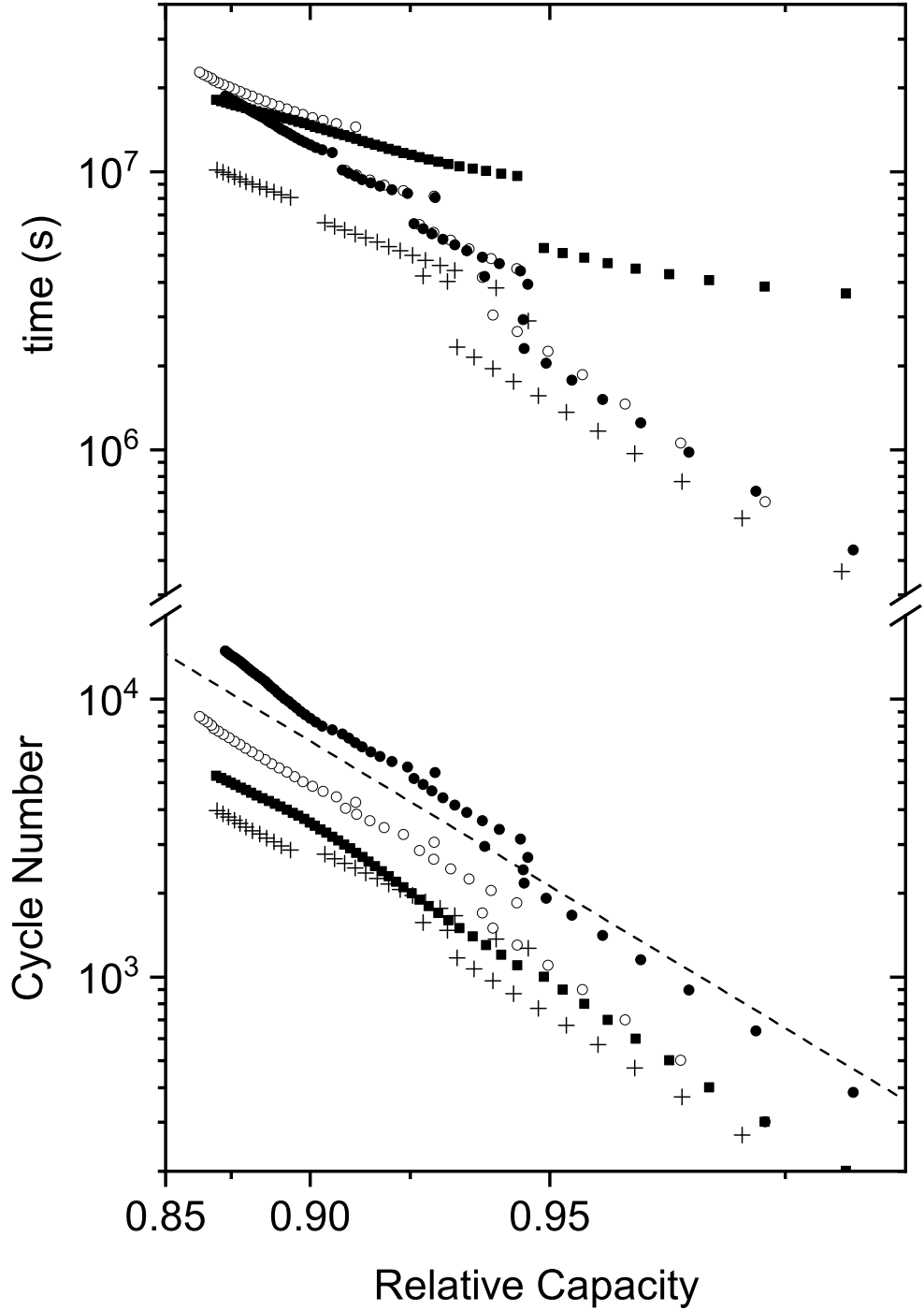}}
\caption{Weibull plot of the ageing curves from four cells of Type 4,  which suffered cycling interruptions. The relative capacity is here used as abscissa in order to represent time, and number of cycles in the same plot.  A dashed line is shown for comparison with the $-0.6$ slope of the SE.  Despite cycling interruptions, the degradation lines exhibit the same slope when plotted with respect to the cycle number.}\label{paused}
\end{figure}

To further support Eq.(\ref{SErel}) we present this last example, which consists of four more ageing trajectories of cells from yet another manufacturer (Type 4). The reason to present these experiments in a separate section is that the cells were aged in interrupted intervals.  The planned ageing programs are described in Table \ref{SettingsCell4}. However, the laboratory was shut down in several occasions during the experiments and therefore, the cells were not aged according to plan. There were intervals of regular cycling  separated by periods of rest, with suboptimal control of the battery state. Type 4 cells  tend to exhibit  negligible ageing during storage, but they do age significantly upon cycling. As such, the decay rate in the interrupted experiments varies from  a standard value during cycling to approximately zero when the test is paused.

In the upper half of FIG. \ref{paused}, one sees that the capacity fade  does not follow a Weibull straight line, when plotted with respect to time (measured from the beginning of the experiment). One also observes vertical jumps  corresponding to time passing without cycles. However, in terms of the cycle number $N_{Cyc}$, the gaps in time are closed and each $q(N_{Cyc})$ follows approximately a straight line with the same slope $-0.6$. Note that during cycling $N_{Cyc}\propto \Delta t$.   Therefore, the capacity fade in terms of time intervals is well represented by   Eq.(\ref{SErel}).

\section{Capacity fade under arbitrary load profiles}\label{Variabletau}
By arbitrary load profiles we refer to operating conditions in which one or more of the cycling parameters change with time. This can mean for instance, changes in the cell temperature, in the electric current, or variations in the average state of charge and cycle depth. Under an arbitrary load profile, a battery will spend time intervals in different conditions, each of which is  characterised by a $\tau$ value. Therefore, $\tau$ becomes time dependent and the model should be generalised to compute such cases.

Eq.(\ref{SErel}) is of great practical use for several reasons. (i) It describes the capacity fade under all regular conditions. (ii) It only depends on the  ageing conditions during the studied time interval, but not on  the previous history. (iii) It contains a single adjustable parameter $\tau$ that can be determined with reasonable accuracy for each case. That Eq.(\ref{SErel})  does not depend explicitly on $t_c^{}$ is essential. The latter is practically impossible to determine from curve fittings, due to the difference in time scales $t_c\ll \tau$ and the large statistical fluctuations  in capacity measurements. However, Eq.(\ref{SErel}) is only accurate if $\tau$ takes approximately constant values over long time-intervals. For a case with variable cycling conditions, the product of terms in the form of Eq.(\ref{SErel}), $q^{}_{\text{lb}}=\prod_{i}\exp\left(-\left[\Delta t_i/\tau_i\right]^{\beta}\right)$ each with the corresponding $\tau_i$ and small time interval $\Delta t_i$, would strongly underestimate the remaining capacity. Instead, a version of Eq.(\ref{SErel}) which  remains accurate for smaller (differential)  time intervals is required. A generalisation of Eq.(\ref{longdT}) in the form

\begin{equation}\label{last}
q^{}_{\text lb}(t)\equiv\exp\left\{-\beta\int\displaylimits^{t}_{0}\frac{du}{u^{1-\beta}\left[\tau(u)\right]^{\beta}}\right\}\,
\end{equation}

\noindent is also not appropriate because the integrable singularity $u^{\beta-1}$ at $u=0$ causes numerical instabilities. The problem is resolved if, instead of starting from  a lower bound, we rewrite the exact relation in Eq.(\ref{generalizedrel})  as

\begin{equation}
\label{reltc}
q(t,t_0)=\exp\left\{-\beta\int\displaylimits^{\Delta t}_{0}\frac{du}{\left[t^{}_c(u)+t_0^{}+u\right]^{1-\beta}\left[\tau(u)\right]^{\beta}}\right\} \,
\end{equation}
\noindent and take into account that $t_0^{}\gg t_c$.  Eq.(\ref{reltc}) has no singularities and its dependence on $t_c^{}$  can be discounted,  taking for $t^{}_c(u)+t_0^{}\approx t_0^{}$ the age of the cell measured from the production date.  The final result is
\begin{equation}
\label{qrelfinal}
q(t,t_0^{})\approx\exp\left\{-\beta\int\displaylimits^{\Delta t}_{0}\frac{du}{\left[t_0^{}+u\right]^{1-\beta}\left[\tau(u)\right]^{\beta}}\right\} \,,
\end{equation}

\noindent which provides $q(t,t_0)$ only using the  $\tau${\scriptsize s} obtained from experiments under various regular conditions, and an estimate of the cell age, $t_0^{}$. Note that any value given to $t_0^{}$, smaller than the actual age of the system, will result in a lower bound to Eq.(\ref{qrelfinal}) that is  more accurate than Eq.(\ref{last}), and Eq.(\ref{SErel}). The relative uncertainty in $q(t,t_0^{})$ due to errors in $t_0^{}$ satisfies
\begin{equation}\label{var}
\frac{|\delta q|}{q}<\beta \left(\frac{t_0^{}}{\text{Min}[\tau]}\right)^{\beta}\left(1-\frac{1}{(1+\Delta t/t_0^{})^{1-\beta}}\right)\frac{|\delta t_0^{}|}{t_0^{}}\,.
\end{equation}

\noindent Taking into account that for life time predictions $\Delta t\sim 0.1 \text{Min}[\tau]$, we can show that $\frac{|\delta q|}{q}< 0.04 \frac{|\delta t_0^{}|}{t_0^{}}$. That is, uncertainties in the age of the battery have small effects in the capacity fade predictions.

\section{Conclusions and outlook}\label{Summary}

We have shown that the capacity fade of at least four different Li-ion cell types follow the same functional form, for all tested conditions. This includes at different temperatures, with or  without cycling.  We have also shown that the diffusion-to-traps model explains this uniformity, with perfect agreement between the experimentally found exponent $\beta=0.6$, and the theoretical prediction $\beta=d/(d+2)$ for $d=3$ dimensions. The decay function is the same stretched exponential that governs a wide variety of relaxation phenomena in natural and social sciences.

The strength of the model for battery applications is further supported by the fact that a single adjustable parameter $\tau$, suffices to describe four cells types under many different conditions. To our knowledge, models found in the literature contain more parameters and have been tested with less experiments, giving often worse fitting quality\cite{SEImodelingreview,Sqrt,Aachen}. One should bear in mind that having a single scaling parameter not only simplifies the modeling; it also guarantees the robustness of the fits. The model with square-root losses, which also contains a single parameter, gives clearly poorer results.

Within the diffusion-to-traps model, the functional form of the capacity fade derives from the  scaling properties of diffusion, and the stochastic character of ions finding traps randomly distributed in space. Capacity fade  is a diffusion-limited process, whichever the trapping mechanism is. This is often overlooked in other   models \cite{SEImodelingreview,Sqrt,Aachen}. Electrochemical models generally focus on reaction kinetics determined by mean-value properties such as leakage or cycling  currents, ignoring the thermal fluctuations which are essential to ion dynamics.  Li-ions hop randomly from site to site, whether there is a net current or not. This hopping (fluctuation) process combined with the random distribution of traps determines a universal form of the capacity decay. Chemical reaction times are commonly  negligible in comparison with diffusion times. Since most ions have to move in order to find a trap, it is the diffusion process which determines the long-time functional form of the capacity fade.

According to the presented evidence, the difference between calendar and cycling ageing seems to come mainly from the amount of movement of the ions, resulting in a different time $\tau$ required for an ion to fall in a trap. There is a single function describing the capacity fade, whether the cell is cycled or not.

The scaling of the mean-square-displacement of ions with time  is always linear in a diffusive transport. Therefore, it is  not affected by cell chemistry. We may then expect that Eq.(\ref{SER}) describes the capacity fade of many cell types. Three of the cell types analysed in this work have  graphite-based anodes and the fourth type, a mix of graphite and LTO. Each of them has its own NMC mix in the cathode. It would be of interest to continue the study with  other cell chemistries.

Cell-specific parameters that govern the ageing are all contained inside a single quantity $\tau$. A complete ageing model for a battery type is therefore built in two steps. In the first step, the $\tau${\scriptsize s} corresponding to several ageing conditions are determined from fittings with Eq.(\ref{SER}). In the second step, an equation must be found that describes the dependence of $\tau$ on the ageing conditions. This equation will contain the cell-type specific parameters. Through the dependence of $\tau$ on ageing conditions, $\tau$ becomes time dependent for cell ageing with arbitrary load profiles. Fade forecast can be made for arbitrary load profiles inserting the time-dependent $\tau$ in Eq.(\ref{qrelfinal}). Examples of such models will be presented in more specific publications in the future.

\section{Methods}\label{Meth}
\subsection{Measurement equipment and settings}\label{Exper}

We have tested four different commercially available cells from four different manufacturers under different aging conditions. Tables below give an overview of the tested parameter sets. The cells are used in applications like photo-voltaic (PV) storage, uninterrupted power systems (UPS) and electric buses. Whether the cells were produced in the same lot, and in the same factory is not known.

\begin{table}[htb]
\caption{Test programs for cell-type 1: Cylindrical, NMC cathode vs graphite anode,  nominal capacity of 2.2 Ah.}\label{SettingsCell1}
\begin{tabular*}{\linewidth}{@{\extracolsep{\fill}}rccc@{}}
\toprule
Test N$^{\text{o}}$ & $T$ ($^{\text{o}}$C)  & C-rate dis-/charge & SOC range ({\%})\\
\midrule
1&60&0&100\\
2&45&0&100\\
3&25&0.2&10--90\\
4&25&0.5&10--90\\
5&45&0.2&10--90\\
6&45&0.5&10--90\\
7&45&0.8&10--90\\
8&45&0.5&10--90\\
9&45&0.5&30-70\\
10&45&0.5&10-50\\
11&45&0.5&50-90\\
\bottomrule
\end{tabular*}
\end{table}

\begin{table}[htb]
\caption{Test programs for cell-type 2: Cylindrical, NMC cathode vs graphite anode,  nominal capacity of 3.4 Ah.}\label{SettingsCell2}
\begin{tabular*}{\linewidth}{@{\extracolsep{\fill}}rccc@{}}
\toprule
Test N$^{\text{o}}$ & $T$ ($^{\text{o}}$C)  & C-rate dis-/charge & SOC range ({\%})\\
\midrule
1&45&0.5/0.2&0--80\\
2&45&0.5/0.2&20--100\\
3&45&0.5/0.2&0--100\\
4&25&1/0.2&0--80\\
5&25&0.5/0.2&0--80\\
\bottomrule
\end{tabular*}
\end{table}

\begin{table}[htb]
\caption{Test programs for cell-type 3: Prismatic, NMC cathode vs graphite anode,  nominal capacity of 50 Ah.}\label{SettingsCell3}
\begin{tabular*}{\linewidth}{@{\extracolsep{\fill}}rccc@{}}
\toprule
Test N$^{\text{o}}$ & $T$ ($^{\text{o}}$C)  & C-rate dis-/charge & SOC range ({\%})\\
\midrule
1&40&4/1&0--90\\
2&40&4/1&40--90\\
3&40&4/1&80--90\\
4&40&6/1&80--90\\
5&25&4/1&40--90\\
6&25&6/1&80--90\\
7&25&0&100\\
8&25&0&60\\
9&25&0&20\\
10&45&0&100\\
11&45&0&60\\
12&45&0&20\\
\bottomrule
\end{tabular*}
\end{table}

\begin{table}[htb]
\caption{Test programs for cell-type 4: Pouch, NMC cathode vs graphite-LTO mix anode,  nominal capacity of 49 Ah.}\label{SettingsCell4}
\begin{tabular*}{\linewidth}{@{\extracolsep{\fill}}rccc@{}}
\toprule
Test N$^{\text{o}}$ & $T$ ($^{\text{o}}$C)  & C-rate dis-/charge & SOC range ({\%})\\
\midrule
1&25&4/3&10--90\\
2&40&4/3&10--90\\
3&40&4&70--90\\
4&40&4&50--90\\
\bottomrule
\end{tabular*}
\end{table}

All tests are started with a reference performance test (RPT) which includes a full discharge and full charge with nominal current. During the RPT the cells are charged and discharged at the upper and lower voltage limit with a constant voltage until the current drops below $1/10$ of the nominal current. In order to monitor capacity fade, RPTs are performed periodically at intervals of not more than $100$ full equivalent cycles, or two weeks for experiments with only  calendar ageing. The measured charge capacity is used as the new actual capacity for the control of depth of discharge (cycle amplitude) in the following cycles.

After each RPT, the cell is discharged to the upper state of charge limit of the corresponding test, before the actual cycling starts. The cycling amplitude is controlled by the Ah throughput given by the actual capacity multiplied with the corresponding depth of discharge. A $10$ s idle time is introduced after each discharge and charge. Hence, the tests are $100$ {\%} continuous cycle ageing with negligible idle calendar ageing. The C-rates are in some cases asymmetric between charge and discharge, because the application is asymmetric, or because the cell does not allow higher charging current. The temperatures given in the tables correspond to values measured by the control system of the temperature chamber. The actual temperature inside the cell can be up to $10$ K higher than the nominal chamber temperature, especially for tests with the highest C-rate.

The ageing tests were performed over a period of typically six months. Within this time the capacity of cells with high temperatures and  wide cycles decreased to about $80$ {\%}.

All cycle aging tests were performed with an ACT0550 cell tester from PEC. The cylindrical cells were mounted in commercial cell holders featuring a point measurement and good air circulation. The prismatic cells were mounted in self-made cell holders with limited air circulation and no heat spreaders. The pouch cells were mounted in self-made cell holders with heat spreaders. No clamping force was applied to any cell. The cells were tempered with two Binder MK720 temperature chambers.
\subsection{Derivation of the asymptotic form of the survival probability}\label{Theor}
The goal is to find the asymptotic long-time dependence of the survival probability of ions diffusing in a  $d$-dimensional space where traps are randomly distributed with average density $n_S$. The general diffusion equation  for the survival probability density $\rho(x,t|S)$ is Eq.(\ref{General}). From $\rho(x,t|S)$, the global survival probability $q(t)$ is derived as described at the beginning of  section \ref{Physics}.

We look at the evolution of an ion that at time $t=0$ is at an arbitrary point and (without losing generality) we choose this point as origin of coordinates. One defines an auxiliary problem, which only differs from the original one in having a  perfectly trapping spherical surface of large radius $R$ centered at the initial position of the ion, in addition to all the common traps. The probability density $\rho'(x,t|S)$  resulting from this new problem satisfies $\rho'(x,t|S) \leq \rho(x,t|S)$ everywhere, and for any $S$, because the addition of extra traps can never increase the survival probability density. As a result, it derives from Eq.(\ref{avdens}) that

\begin{equation}
\bar{\rho}(x,t)\geq\sum_S\rho'(x,t|S)P(S)\, .\label{firstbound}
\end{equation}
\noindent Since the r.h.s. of Eq.(\ref{firstbound}) is a sum of non-negative terms, it is larger or equal than any of its partial sums. Thus
\begin{equation}
\bar{\rho}(x,t)\geq \rho_0'(x,t)P_0\, ,\label{secondbound}
\end{equation}

\noindent where $P_0$ is the probability that none of the traps lies inside the sphere, and $ \rho_0'(x,t)$ is the corresponding $\rho'(x,t|S)$. Note that due to the absorbing spherical surface, all configurations with no trap in the sphere lead to the same solution $\rho'(x,t|S)$, which we just called $ \rho_0'(x,t)$. The usefulness of this trick relies on our ability to calculate $\rho_0'(x,t)$, and $P_0$ allowing us to find a lower bound to $q(t)$. For randomly distributed traps, it follows from the Poisson distribution that
\begin{equation}
P_0=\exp(-n_Sc_d(R+a)^d)\, ,\label{P0}
\end{equation}
\noindent where $c_d=\frac{\pi^{d/2}}{\Gamma(d/2+1)}$ and $a$ the action reach of a trap. The solution  for $\rho_0'(x,t)$  is
\begin{equation}
\rho_0'(x,t)=\sum_i\phi_i(0)\phi_i(x)\exp(-\lambda_i Dt)\, , \label{FS}
\end{equation}
\noindent where $\phi_i(x)$ and $\lambda_i$ are the eigenfunctions and eigenvalues, respectively, of the partial differential equation  $\nabla^2\phi_i(x)+\lambda_i\phi_i(x)=0$, with  boundary condition $\phi_i(R)=0$, and $|\phi_i(0)|<\infty$ \cite{Tikhonov1990}. For any $d$, the eigenvalues are $\lambda_i=\gamma_i/R^2$, with $\gamma_i$ dimensionless numbers, independent of $R$.  In three dimensions ($d=3$), $\gamma_i=i\pi$, $\forall i\in\mathbb{N}^* $. The long-time behavior of $\rho_0'(x,t)$  is dominated by the smallest eigenvalue $\lambda_1$; thus
 \begin{equation}
\rho_0'(x,t)\approx  \phi_1(0)\phi_1(x)\exp(-\gamma_1 Dt/R^2)\; .\label{lambda0}
\end{equation}
Substituting Eq.(\ref{P0}), and  Eq.(\ref{lambda0}) in  Eq.(\ref{secondbound}) and integrating over the volume we obtain
\begin{equation}
q(t)\geq q(R,t)\equiv A_d\exp(-\gamma_1 Dt/R^2-n_Sc_d(R+a)^d)\, ,\label{finalbound}
\end{equation}

\noindent where $ A_d=\phi_1(0)\int d^dx\phi_1(x)$ is a number that depends on $d$, but not on $R$. Since the choice of $R$ is arbitrary, the best lower-bound to $q(t)$ is obtained with the $R$ that minimises the exponential argument. The minimising $R$ is

\begin{equation}
R_0^{}\approx\left(\frac{2\gamma_1D}{dc_dn_S}t\right)^{\frac{1}{d+2}}\, ,\label{MSD}
\end{equation}
\noindent and therefore the best lower bound to $q(t)$ is
\begin{equation}
q(t)\geq A_d \exp(-[t/\tau]^{\beta})\,,\label{SERm}
\end{equation}
\noindent with $\beta$, and $\tau^{-1}$ given by Eq.(\ref{be}), and Eq.(\ref{tau}), respectively. It was taken into account that around the maximum of $q(R,t)$,  $a/R\ll 1$

\begin{figure}[htb]
\centerline{\includegraphics[width=0.8\linewidth]{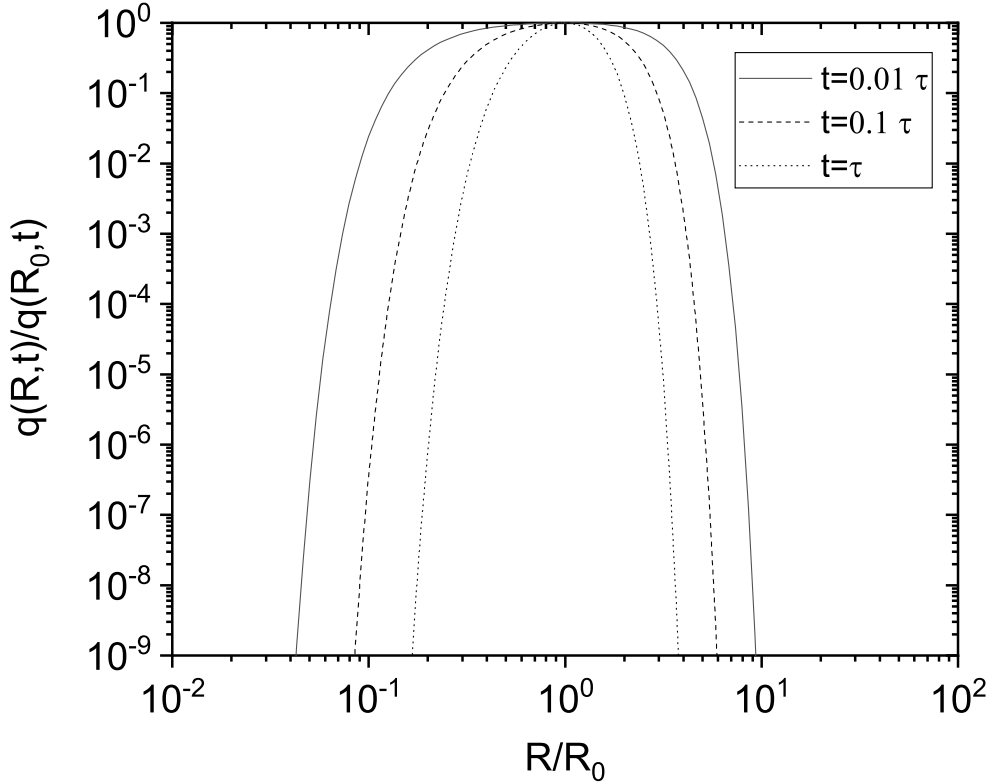}}
\caption{Dependence of the lower bound to the survival probability on $R$, for different times. The strongly peaked function, and its narrowing down as the time increases, indicates the statistical significance of the typical distance $R_0^{}$ and the asymptotic (long-time) robustness of the estimated lower bound.}\label{Rsignificance}
\end{figure}
The statistical significance of $R_0(t)$ is shown in FIG. \ref{Rsignificance}. The narrow peak of the product probability at $R_0$   becomes even narrower at longer times, indicating that the extreme estimate is asymptotically precise.

\section{Supporting information}\label{Support}

\begin{figure}[htb]
\includegraphics[width=0.8\linewidth]{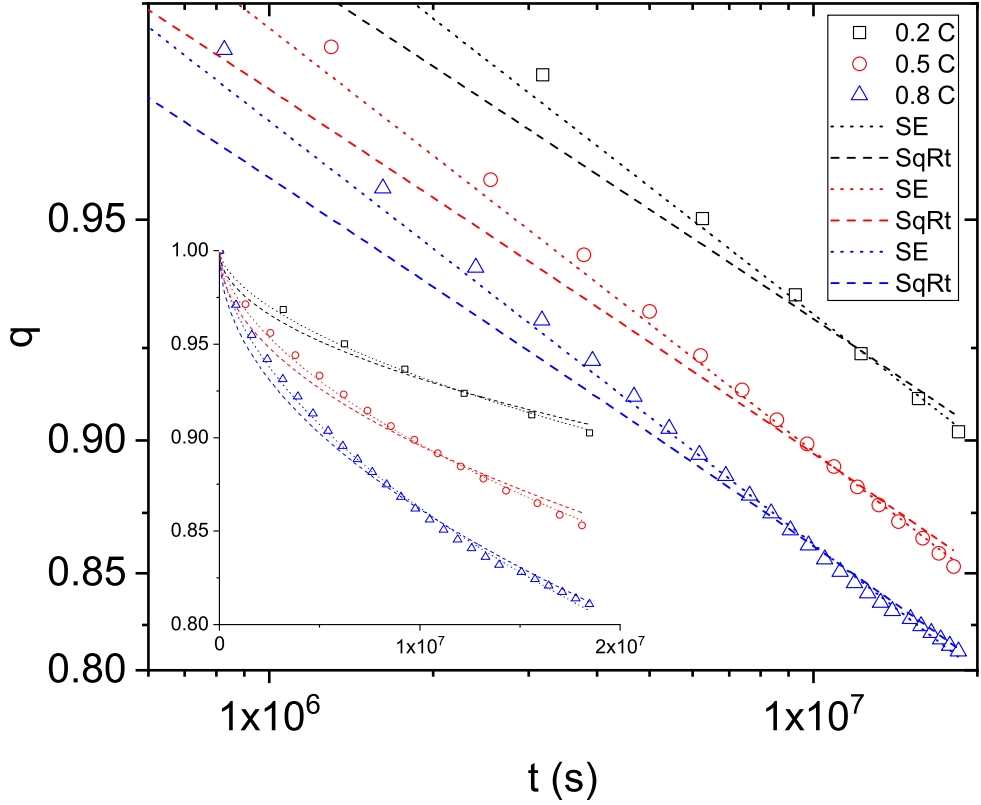}
\caption{Comparison of model fittings for three different capacity fade trajectories using  $y=\exp(-[t/\tau]^{0.6})$ (dotted lines), and $q=1-\sqrt{t/\tau}$ (dashed lines). The main panel shows the axes with Weibull scaling; the inset displays the same data in linear scale. Note the overall incorrect curvature (and slope in the Weibull plot) of the $q=1-\sqrt{t/\tau}$ model, which does not allow to predict the long-time behavior from earlier decay data. On other hand, $y=\exp(-[t/\tau]^{0.6})$ reproduces the data much better over the whole time interval.}
\label{fig:Meth1}
\end{figure}

In FIG. \ref{fig:Meth1} we show the capacity retention of three Type 1 cells aged at different cycling rates, together with the best fitting curves using Eq.(\ref{SER}), and the common square-root model. The fitting quality with Eq.(\ref{SER}) is clearly better. The square-root model does not have the correct curvature.

%%%%%%%%%%%%%%%%%%%%%%%%%%%%%%%%%%%%%%%%%%%%%%%%%%%%%%%%%%%%%%%%%%%%
% Acknowledgments
%%%%%%%%%%%%%%%%%%%%%%%%%%%%%%%%%%%%%%%%%%%%%%%%%%%%%%%%%%%%%%%%%%%%
\begin{acknowledgments}

The authors thank Thomas Christen, Joerg Lehmann, Kay Henken, Sophie Vanderspar, and Elise Fahy for valuable discussions and  for kindly proofreading the article.

\end{acknowledgments}
%%%%%%%%%%%%%%%%%%%%%%%%%%%%%%%%%%%%%%%%%%%%%%%%%%%%%%%%%%%%%%%%%%%%
%\appendix
%\section{Integration of ... }\label{FDTApp}
%%%%%%%%%%%%%%%%%%%%%%%%%%%%%%%%%%%%%%%%%%%%%%%%%%%%%%%%%%%%%%%%%%%%
% Include bibtex file here with references
%%%%%%%%%%%%%%%%%%%%%%%%%%%%%%%%%%%%%%%%%%%%%%%%%%%%%%%%%%%%%%%%%%%%
%\bibliography{refs}{}
%

\end{document}